\documentclass[twocolumn,aps,prl,groupedaddress,showpacs, superscriptaddress]{revtex4-1}
\usepackage{amsfonts}
\usepackage{amsmath}
\usepackage{amssymb}
\usepackage{txfonts}
\usepackage{pxfonts}
\usepackage{graphicx,bm,units,yfonts}
\usepackage[table]{xcolor}
\usepackage{hyperref}

\newcommand{\RM}{{\mathbb R}}
\newcommand{\SM}{{\mathbb S}}
\newcommand{\TM}{{\mathbb T}}
\newcommand{\ZM}{{\mathbb Z}}
\newcommand{\KM}{{\mathbb K}}

\newcommand{\Aa}{{\mathcal A}}

\newcommand{\Tt}{{\mathcal T}}

\begin{document}

\title{Observation of topological edge modes in a quasi-periodic acoustic waveguide}

\author{David J. Apigo}
\affiliation{Department of Physics, New Jersey Institute of Technology, Newark, NJ, USA}

\author{Wenting Cheng}
\affiliation{Department of Physics, New Jersey Institute of Technology, Newark, NJ, USA}

\author{Kyle F. Dobiszewski}
\affiliation{Albert Dormans Honors College, New Jesrey Institute of Technology, Newark, NJ, USA}

\author{Emil Prodan}
\affiliation{Department of Physics, Yeshiva University, New York, NY, USA}

\author{Camelia Prodan}
\affiliation{Department of Physics, New Jersey Institute of Technology, Newark, NJ, USA}

\begin{abstract}
Topological boundary and interface modes are generated in an acoustic waveguide by simple quasi-periodic patterning of the walls. The procedure opens many topological gaps in the resonant spectrum and qualitative as well as quantitative assessments of their topological character are supplied. In particular, computations of the bulk invariant for the continuum wave equation are performed. The experimental measurements reproduce the theoretical predictions with high fidelity. In particular, acoustic modes with high Q-factors localized in the middle of a breathable waveguide are engineered by a simple patterning of the walls.
\end{abstract}

\maketitle

The ideas based on topological concepts \cite{ThoulessPRL1982,HaldanePRL1988} have revolutionized the field of condensed matter physics and led to the discovery of topological insulators and superconductors. The latter have been classified at the end of the previous decade \cite{SRFL2008,QiPRB2008,Kit2009,RSFL2010} and
a table of strong topological phases has been conjectured. One of their common characteristics is the emergence of disorder-immune boundary modes whenever a sample is halved. Physics akin to that of topological condensed matter systems has been also predicted in classical wave-supporting materials \cite{HaldaneRaghu2008,PP2009} and many examples of topological metamaterials have been reported in the literature \cite{WangNature2009,NashPNAS2015,HafeziNatPhot2013,WuPRL2015,
SusstrunkScience2015,KaneNatPhys2013,PauloseNatPhys2015,
ProdanNatComm2017,KhanikaevNatPhot2017,MousaviNatComm2015,
RuzzeneArxiv2017,ChaunsaliPRB2018,ChernArxiv2018,PalJAP2016}. 

At the same time, it has been pointed out that the periodic table of topological systems is highly enhanced if more complex systems are considered, such as the quasi-periodic or quasi-crystalline ones \cite{KLR2012,VZK2013,ProdanPRB2015,BabouxPRB2017}. In \cite{Apigo2}, $K$-theoretic arguments \cite{Bel86,PS} were applied for quasi-periodically coupled discrete mechanical resonators. The finding was that, if these are single-mode resonators, then every gap in the bulk resonant spectrum is topological, in the sense that it will be completely filled by boundary spectrum under any boundary condition. The practical value of the finding is that the quasi-periodic Hamiltonians display a large number of topological gaps, hence one can generate localized wave-modes in both space and energy by simply halving the system.

In this work, we put these general principles to the test in a completely different regime and we implement them for the first time using sound waves. Acoustic setups have been successfully used in the past to generate topological edge modes \cite{He, Xiao,NiArxiv2018} and even to map the Hofstadter butterfly \cite{RichouxEPL2002}. In particular, \cite{NiArxiv2018,RichouxEPL2002} introduced re-configurable acoustic resonant structures where the building blocks are sealed acoustic chambers connected via thin bridges. They have isolated resonant modes, hence these structures fall under the umbrella of patterned resonators introduced in \cite{Apigo2} and they can be analyzed by similar methods. However, these types of acoustic structures are not breathable, which is a key requirement for many practical applications. As such, here we ask the question: Can one generate topological edge and interface modes by patterning the walls of an acoustic waveguide without impeding the air flow?

As we shall see, the answer is yes, but the methods of analysis are very different from those introduced in \cite{Apigo2}. Indeed, the picture of coupled discrete resonators is no longer applicable and a full continuum medium treatment must be employed for the theoretical analysis. Furthermore, the topological character of the spectral gaps cannot be taken for granted because the waveguide supports many overlapping modes. As such, a new assessment of the topological character is introduced based on the continuum version of the lattice non-commutative Chern number proposed in \cite{ProdanPRB2015}, achieved in \cite{BourneMPAG2018}. This invariant is here evaluated numerically using the methods developed in \cite{ProdanSpringer2017,ProdanJPA2018}. Let us recall from \cite{Apigo2} that the role of aperiodicity in this type of applications is to generate virtual dimensions and, as we shall see \cite{Suppl}, the Chern number mentioned above is defined on a 3-dimensional non-commutative manifold, while for discrete patterns is on a 2-dimensional manifold.  

At the experimental level, challenges exist because some of the spectral bands are very narrow and this, together with the aperiodicity, can lead to irregular mode profiles, although the bulk states are extended. As such, the only way to accurately map the bulk spectrum is to collect data from a large number of points along the waveguide. Following this protocol, we map not only the frequecy but also the spatial profile of the bulk modes. Furthermore, inside the topological bulk gaps, we were able to detect sharp edge modes, which flow with the phason degree of freedom in a manner consistent with the computed Chern numbers.

\begin{figure*}[!ht]
\centering
\includegraphics[width=\linewidth]{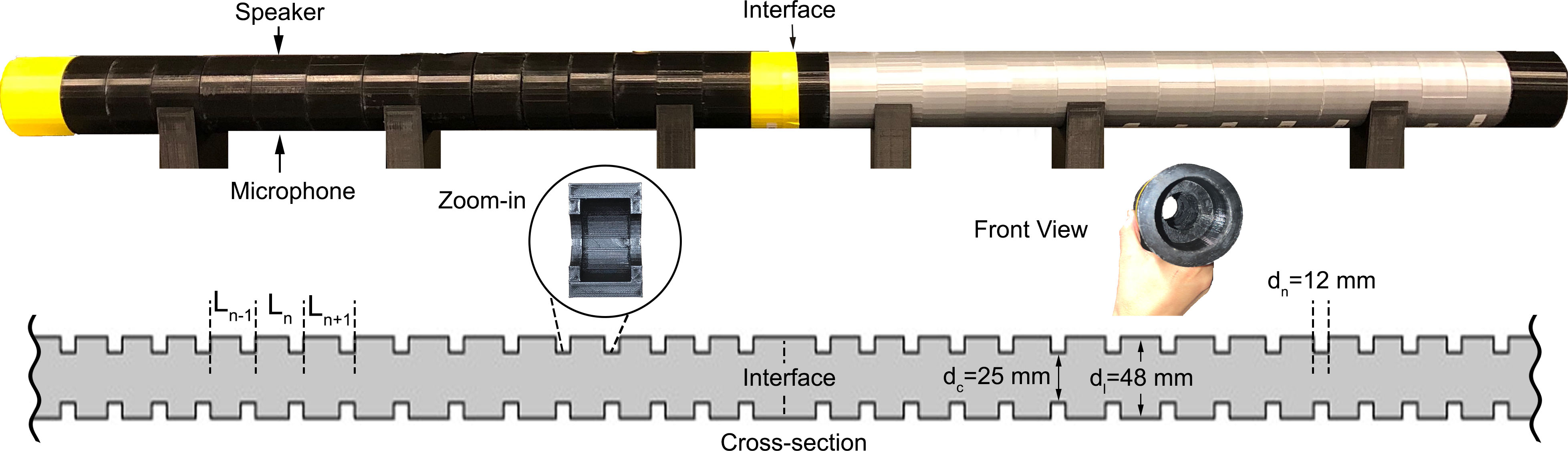}
\caption{\small Top: photograph of the waveguide configuration used to measure topological interface modes. Bottom: Cross-section and geometrical parameters. The waveguide consists of interlocking 3D printed PLA parts as shown in the insert and it is mirrored relative to the domain wall indicated by the dashed line. For experimentation, a speaker is placed at portholes accessible in each chamber and a piezoelectric microphone is inserted into an opposite porthole. The portholes that are not in use are sealed. The lengths $L_n$ were generated with Eq.~\eqref{Eq:Algorithm} and their average was fixed at $L_{\rm avg}=40$ mm. The parameters in Eq.~\eqref{Eq:Algorithm} were fixed at $\Delta L=0.2L_{\rm avg}$ and $\theta = \frac{2 \pi}{\sqrt{117}}$. This particular irrational fraction of $2 \pi$ accepts a good rational approximation $\theta= \frac{9\pi}{48} + \mathcal O(10^{-3})$, which was used in some of the numerical calculations. The system was also run without a domain wall, for bulk and edge measurements.}
\label{Fig:Setup}
\end{figure*}

The quasi-periodic acoustic waveguide consists of a uniform cylindrical tube decorated with walls. The parts were 3D-printed out of polylactic acid (PLA) using an Ultimaker 3 and then assembled as in Fig.~\ref{Fig:Setup}.The walls have identical thickness but the spacings between adjacent walls are modulated according to the algorithm:
\begin{equation}\label{Eq:Algorithm}
L_n = L_{\rm avg} +\Delta L \,\sin(n \theta + \phi ), \quad n \in \ZM.
\end{equation}
The geometric parameters used in the experiments are supplied in Fig.~\ref{Fig:Setup}. To make the above labels meaningful, we assume that the waveguide is centered at a point inside $L_0$. In Equation \eqref{Eq:Algorithm}, $\theta$ is an angle incommensurate with $2\pi$, which will be kept fixed during the measurements, and $\phi$ is the phason, which should be let to vary. For example, a simple relabeling $n \rightarrow n+m$, which corresponds to re-centering the waveguide, will change $\phi$ into $(\phi + m \theta) {\rm mod}\, 2\pi$. Since $\theta$ is incommensurate, these relabelings alone will sample the phason densely in the $[0,2\pi]$ interval. $L_{\rm avg}$ in Equation \eqref{Eq:Algorithm} is the average distance between the walls and $\Delta L$ sets the magnitude of the fluctuations in $L_n$. 

In the inset of Fig.~\ref{Fig:Setup}, we show a front view of the waveguide, confirming that air can flow freely through the structure. It is then somewhat striking that, with the proposed patterning, we can stop sound propagation over several intervals of frequencies and, furthermore, we can generate, very much on demand, topological sound modes localized at any desired location along the tube. As opposed to an ordinary resonant mode produced in a fully sealed acoustic chamber, the interface modes produced in the present work have less contact with the boundary, hence they are expected to have very high Q-factors, a much desired characteristic for practical applications.

\begin{figure}[t]
\centering
\includegraphics[width=\linewidth]{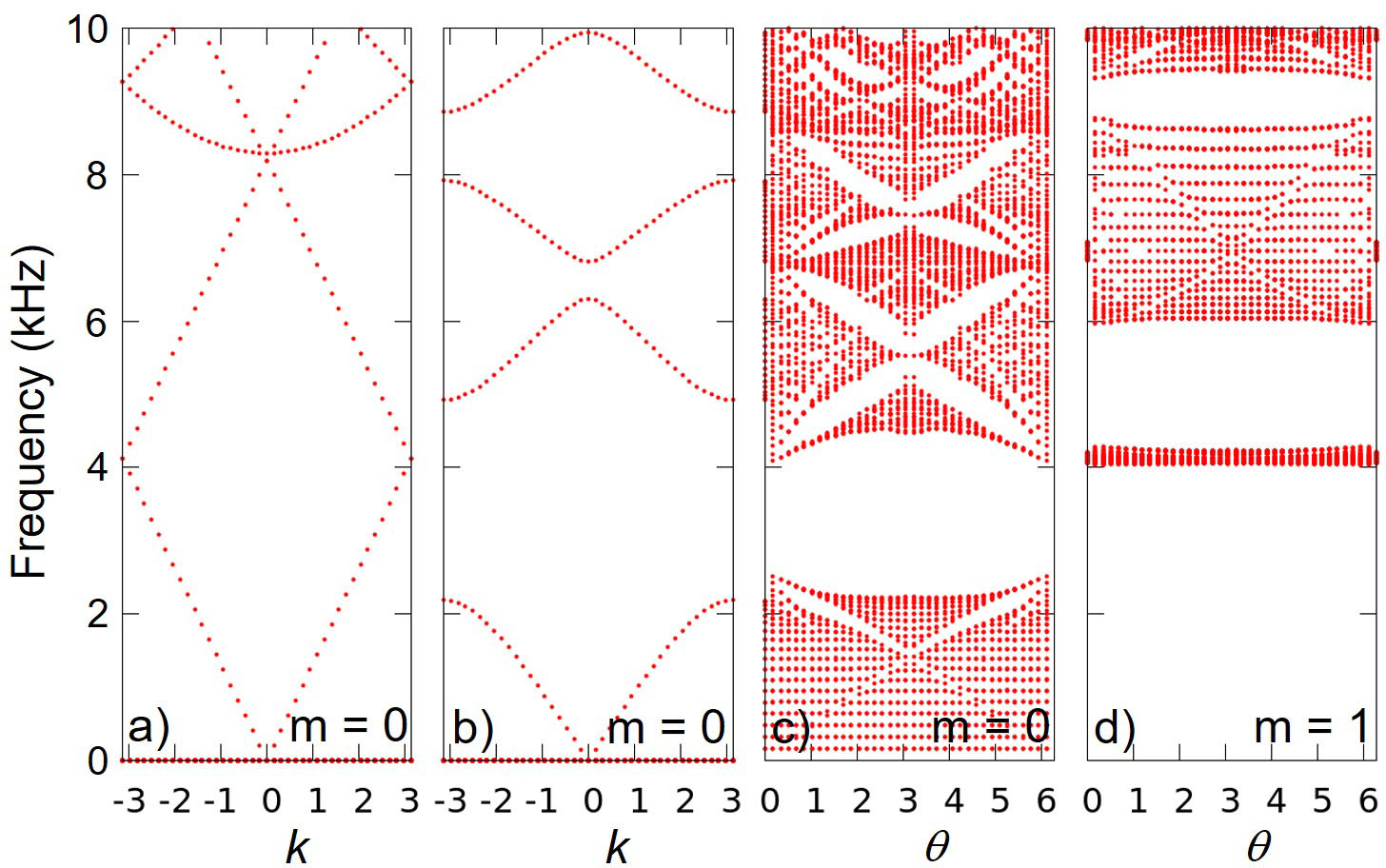}
\caption{\small a) Dispersion of the acoustic modes for the un-patterned waveguide, for $m=0$ sector. b) The band structure of a periodically patterned waveguide ({\it i.e.} $\theta=0$), for $m=0$ sector. c,d) Resonant spectrum of a patterned waveguide as function of $\theta$, for $m=0,1$ sectors, respectively.}
\label{Fig:BandStructure}
\end{figure}

To understand the effect of the patterning, we report in Fig.~\ref{Fig:BandStructure} the dispersion of the acoustic modes for clean and periodically ($L_n = L_{\rm avg}$) patterned waveguides, as well as the resonant spectrum of the aperiodically patterned waveguide ($L_n$ set by \eqref{Eq:Algorithm}). As expected for quasi 1-dimensional wave propagation, the periodic pattern opens spectral gaps in the gapless spectrum of the clean tube. These gaps, however, are not topological. The role of aperiodicity is to open additional gaps in the spectrum that, as one can see, resemble quite closely the Hofstadter butterfly \cite{HofstadterPRB1976}, when mapped as function of $\theta$. As we shall see, these are the gaps that carry non-trivial bulk topological invariants prompting the topological edge and interface modes. Let us mention that the spectra in Fig.~\ref{Fig:BandStructure} were produced with an in-house Fortran code, which diagonalizes the Laplace operator expressed in the cylindrical coordinates $(\rho,z)$ and resolved over the azimuthal symmetry sectors. In appropriate units, the operator reads:
\begin{equation}
\Delta_m = -\frac{1}{\rho} \frac{\partial}{\partial \rho}\rho \frac{\partial }{\partial \rho} + \frac{m^2}{\rho^2}  - \frac{\partial^2 }{\partial z^2}, \quad m=0,\pm 1,\ldots,
\end{equation}
and von Neumann condition is considered at the boundary. Recall that the latter is set by $\theta$ and $\phi$, hence $\Delta_m$ depends in a fundamental way on these parameters. The Laplace operator was discretized using finite differences.

The protocol for acoustic data acquisition was as follows. Sinusoidal signals of duration 1~s and amplitude of 0.5~V were produced by a Rigol DG1022 function generator, amplified by a Crown XLS 2502 power amplifier with the gain set to 6, and then applied on a CUI Inc. GF0501 speaker, placed at one of the portholes. A PCB Piezotronics Model-378C10 microphone and a PCB Piezotronics Model-485B12 power conditioner acquired the acoustic signals at a porthole opposite the speaker (see Fig.~\ref{Fig:Setup}). To account for the frequency-dependent response of the components, a separate measurement is performed with the waveguide removed but speaker and microphone kept in the same positions. All readings are normalized by the output of these measurements. The outputs were read by a custom LabVIEW code via a National Instruments USB-6112 data acquisition box and the ratio of the two measurements is stored on a computer for graphic renderings.

\begin{figure}[t]
\centering
\includegraphics[width=\linewidth]{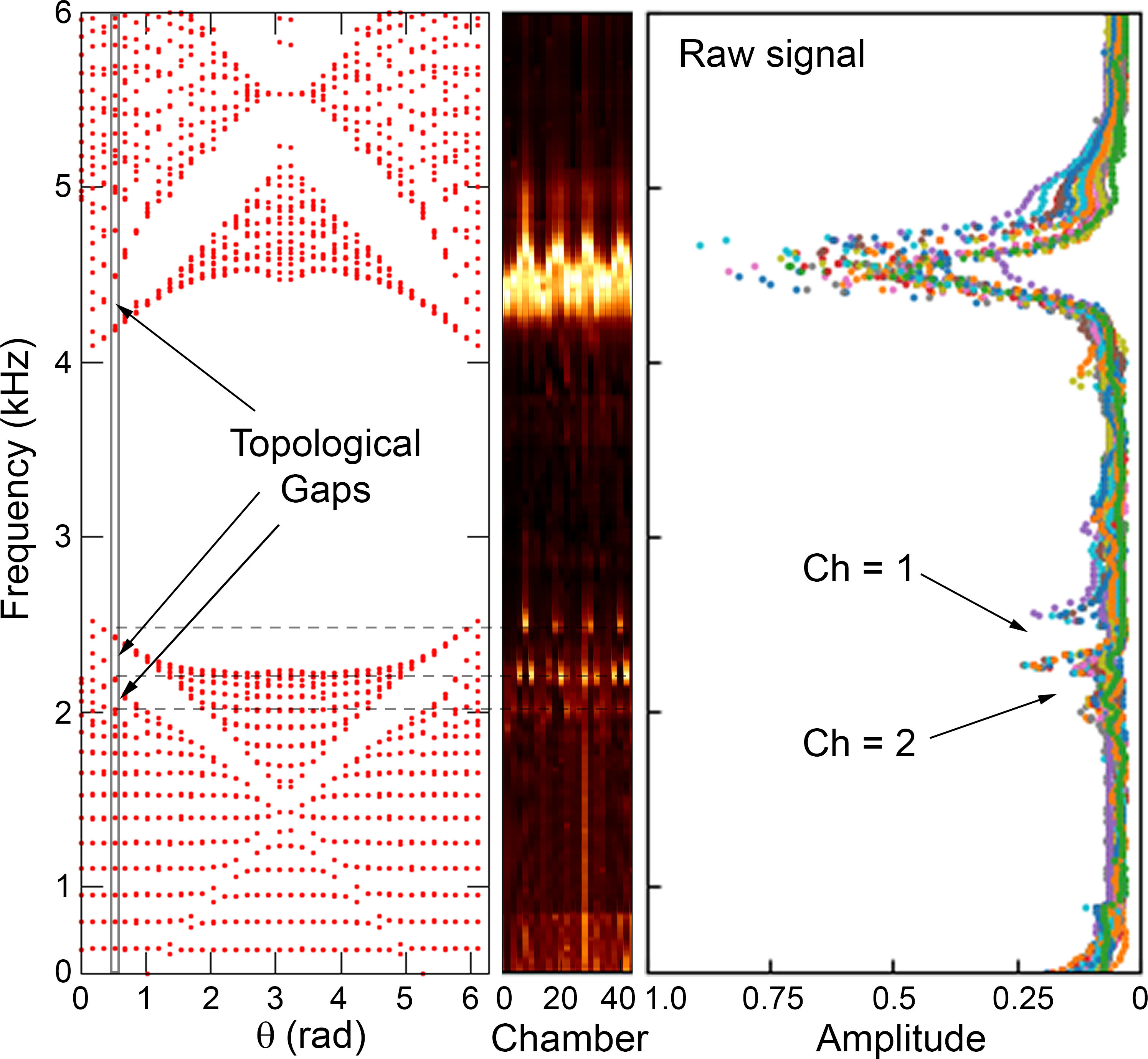}
\caption{\small Bulk resonant spectrum for the geometry described in Fig.~\ref{Fig:Setup}. Left: Theoretical resonant spectrum reproduced from Fig.~\ref{Fig:BandStructure}(c), with arrows indicating the topological gaps. The vertical marking identifies $\theta = \frac{2 \pi}{\sqrt{117}}$, used in experiments. Center: Normalized microphone readings from the center of 48 chambers, recorded over a wide frequency interval. Right: Collapse on the frequency axis of the intensity plot reported in the mid panel. Three spectral gaps can be clearly identified in the experimental data and seen to be well aligned with the theoretical calculations. The values of the Chern numbers for the two topological gaps are also indicated.}
\label{Fig:BulkSpectrum}
\end{figure}

For the bulk measurements, the protocol was repeated for all 48 chambers of a patterned waveguide, with frequency scans from 500 to 6000 Hz in 25 Hz steps. The results are reported in Fig.~\ref{Fig:BulkSpectrum}. When the data is rendered as function of frequency and chamber index, clear extended acoustic modes can be identified. Furthermore, when the data is collapsed on the frequency axis, clear spectral gaps can be identified, two of which are predicted to be topological. Unfortunately, the $m=0,1$ spectra overlap above the non-topological gap (see Fig.~\ref{Fig:BandStructure}) and the higher frequency topological gaps could not experimentally resolved. Let us note that the agreement between experiment and theory in Fig.~\ref{Fig:BulkSpectrum} is less than 5\%.

To assess the topological character of the gaps, we employ the bulk-boundary correspondence for continuum models established in \cite{BourneMPAG2018}. The bulk-topological invariant is supplied by the non-commutative Chern number of the gap projection $P_G = \chi_{(-\infty,G]}\big (\Delta_m(\phi) - G\big )$:
\begin{equation}\label{Eq:BulkInvariant}
{\rm Ch}(P_G) = {\rm Tr}_L\big ( P_G[\partial_\phi P_G,[Z,P_G]] \big ),
\end{equation}
where $Z$ is the position operator parallel to the tube and ${\rm Tr}_L$ is the trace per length. The invariant can be computed at any arbitrary but fixed phason value, which is a consequence of Birkhoff ergodic theorem \cite{Bir}. With the Laplacian discretized on a lattice via finite differences, Eq.~\eqref{Eq:BulkInvariant} was evaluated using methods which are by now standard \cite{ProdanSpringer2017,ProdanJPA2018}. The results are reported in Fig~\ref{Fig:BulkSpectrum}, confirming that the smaller gaps are topological. Furthermore, \cite{BourneMPAG2018} established the existence of a boundary topological invariant which counts the number of chiral boundary bands, as well as the equality between the bulk and boundary invariants.

\begin{figure}[t]
\centering
\includegraphics[width=\linewidth]{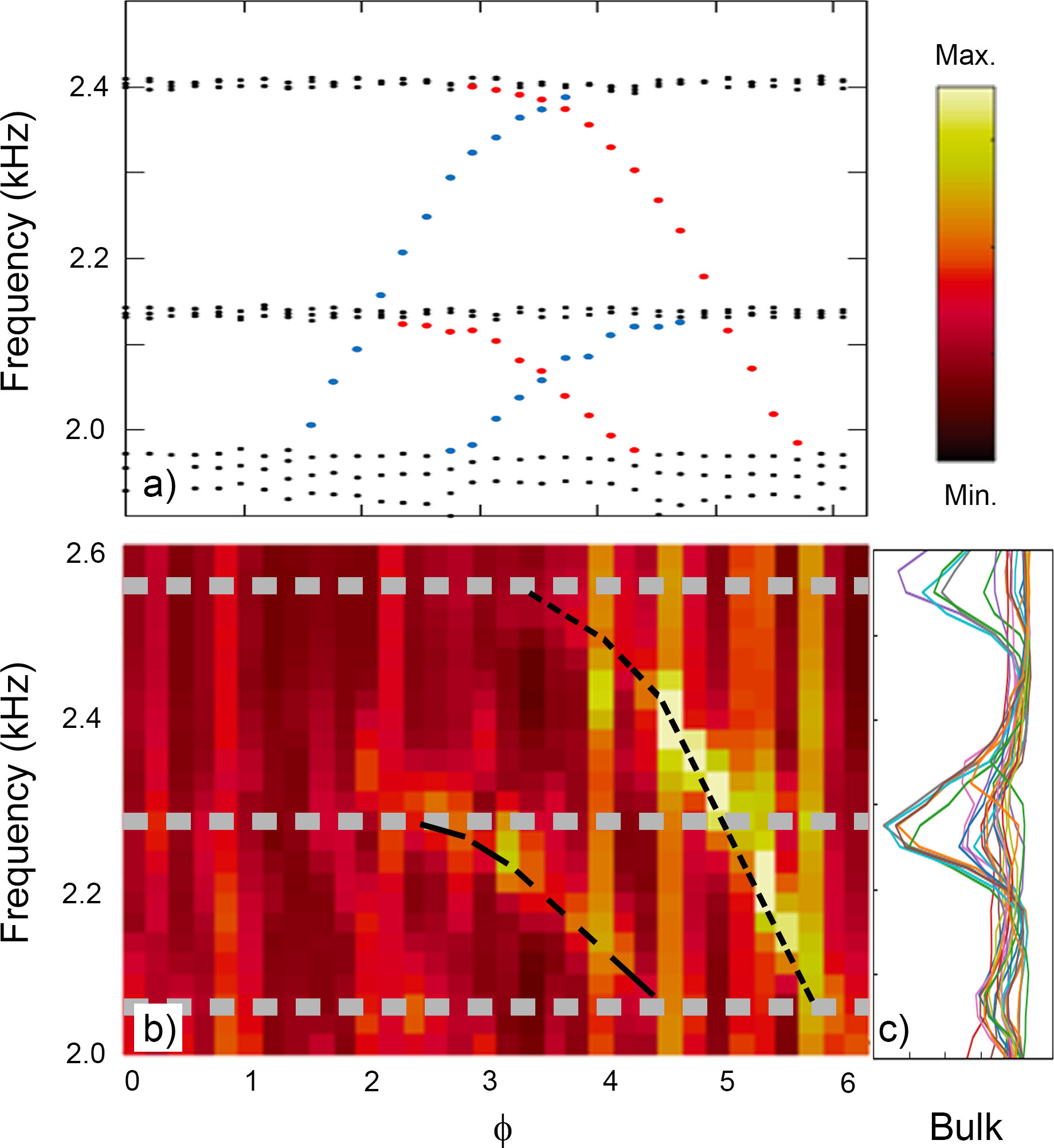}
\caption{\small  Topological edge spectrum. a) Theoretical prediction of the spectral flow against the phason parameter $\phi$, demonstrating the existence of chiral bands. The red/blue marks relate to the left/right edge of the waveguide, respectively. b) Experimental mapping of the spectral flow, confirming the existence of chiral bands. c) The measurements for bulk spectrum, reproduced from Fig.~\ref{Fig:BulkSpectrum}, indicating the position of the bulk gap edges.}
\label{Fig:Edge}
\end{figure}

The presence of chiral modes, in accordance to the above bulk-boundary principle, is confirmed by our numerical simulations reported in  Fig.~\ref{Fig:Edge}a. To map the boundary modes experimentally, the acquisition protocol was applied on the second chamber from the left physical edge, which was plugged. The frequency was swept from 2.0 to 2.6~kHz in steps of 25~Hz and the value of the phason was modified by moving the physical edge sequentially to the right, hence from $L_0$ to $L_n$, $n=1,2,\ldots$. The results are presented in Fig.~\ref{Fig:Edge}b and they indeed confirm the existence of one chiral band in the upper topological gap and two such bands in the lower topological gap. For reference, we reproduced in panel c) the experimental data from Fig.~\ref{Fig:BulkSpectrum}, from where the exact position of the bulk edges can be inferred. As one can see, the boundary resonances occur inside the bulk gaps and the dispersion with $\phi$ is consistent with the theoretical prediction.

We now demonstrate that a localized topological edge mode can be created without the assistance of any plug. For this, we consider a domain wall configuration: 
$$\ldots |L_{31}|L_{30}|L_{29}|L_{29}|L_{30}|L_{31}|\ldots $$ 
where the waveguide is mirror-reflected relative to left edge of $L_{29}$ chamber. This particular index was chosen because moving the origin to that chamber generates a phason $\phi = (29 \theta){\rm mod}\, 2\pi$, which coincides with the value where strong mid-gap edge modes were observed in the first topological gap. Since Eq.~\ref{Eq:BulkInvariant} is odd under reflection, with this patterning, an interface between topological systems with opposite Chern numbers is created As such, the bulk-boundary principle predicts the emergence of $2\times {\rm Ch}$ acoustic modes localized at the interface.

\begin{figure}
\centering
\includegraphics[width=\linewidth]{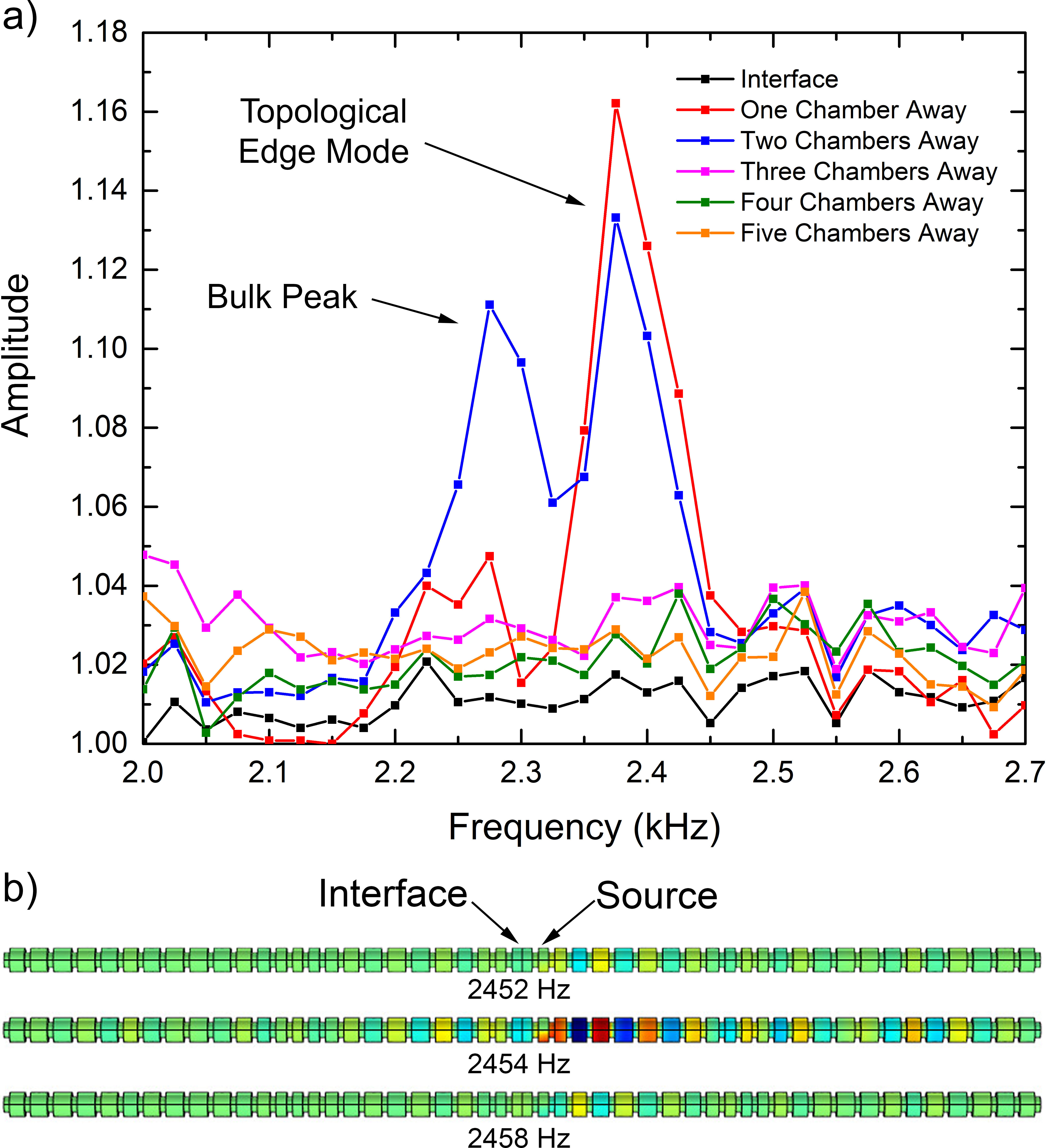}
\caption{\small a) Topological interface mode, measured for a waveguide configuration similar to that in Fig.~\ref{Fig:Setup}. The spatial localization of the interface mode was mapped by moving the speaker and microphone incrementally away from the domain wall. b) The topological interface mode is also observed in COMSOL simulations. Red, blue and green colors represent high, low and zero pressure variations, respectively. }
\label{Fig:Localized_sound}
\end{figure}

The experimental measurements are reported in Fig.~\ref{Fig:Localized_sound}(a). The frequencies were swept as in Fig.~\ref{Fig:Edge} and, in order to probe the localization of the acoustic modes, the speaker and microphone were placed at several portholes at and away from the interface. A strong and sharp resonance was detected in the first topological gap (${\rm Ch}=1$), when the measurements were performed one and two chambers away from the interface. The resonance was not detectable further away from the interface or at the interface itself. A similar resonance can be detected at the other side of the interface, leading to a full confirmation of the topological bulk-boundary prediction. The interface mode is also observed in a standard COMSOL simulation, as shown in Fig.~\ref{Fig:Localized_sound}(b).

In conclusion, we have demonstrated that topological edge and interface modes can be created by a simple quasi-periodic patterning of an acoustic waveguide. The topological gaps can be easily identified when the resonant spectrum is mapped as function of modulation parameter $\theta$. Furthermore, a topological invariant was computed and shown to be in agreement with the number of observed topological chiral edge modes.   

As we have seen, quasi-periodicity opens topological gaps inside the bands of the periodic structure, which resemble the Hofstadter butterfly when mapped as function of $\theta$. Optimization over $\Delta L$ in Eq.~\eqref{Eq:Algorithm} and the geometric parameters of the tube, as well as improvements in materials ({\it e.g.} by replacing the polymer with metal), can highly enhance these topological gaps and the Q-factors of the topological boundary and interface modes. Other than that, the procedure requires no further fine tuning and, due to its simplicity, we believe it can be easily incorporated in practical applications. The present analysis can also serve as a model for acoustic implementations of many other promising aperiodic structures  \cite{ProdanJGP2018}. \\

\acknowledgments{ All authors acknowledge support from the W. M. Keck Foundation.}

\section{Supplemental}

\begin{figure*}[t]
\centering
\includegraphics[width=0.8\linewidth]{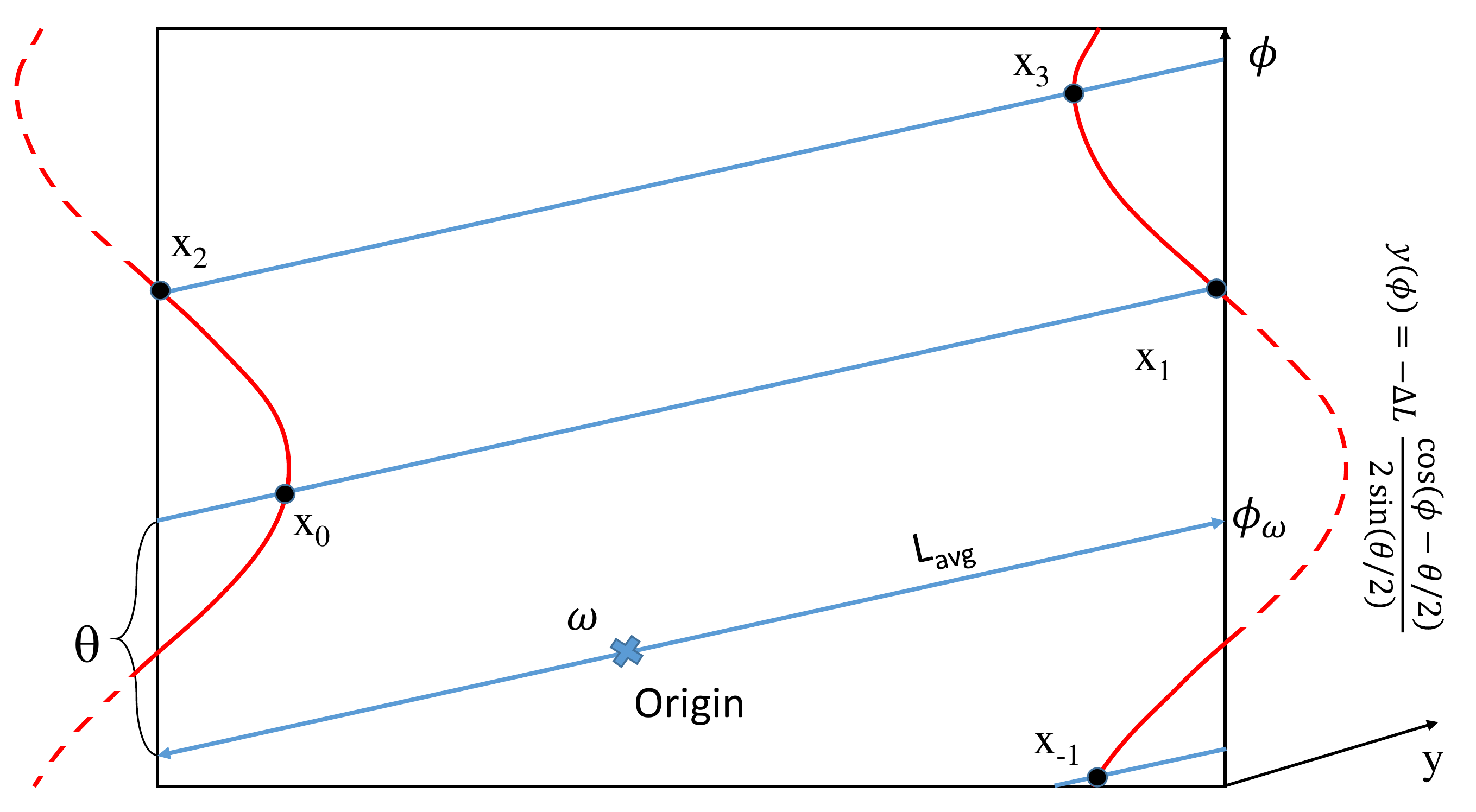}
\caption{\small The virtual manifold associated to the patterned waveguide is a 2-torus.}
\label{Fig:Pattern}
\end{figure*}

The aperiodic continuum systems are quite different from the discrete aperiodic ones for the following reasons:
\begin{itemize}
\item They can be halved at any point of their axis. Hence, the phason $\phi$ alone does not specify completely the configuration of the waveguide, for we also need to know where the origin of the Euclidean space is located relative to the walls of the patterned waveguide. This is so because, by convention, it is at this origin where the cut is made and the edge modes emerge. 
\item The bulk topological invariant, while still a formal Chern number, is defined on a different algebra of observables.
\item The proof \cite{BourneMPAG2018} of the quantization and stability of the topological invariants for continuum models also proceeds quite differently from the one for discrete models \cite{PS}. 
\end{itemize}
The aim of this note is to walk the reader through \cite{BourneMPAG2018}, as adopted to the acoustic waveguide analyzed in the main text.

\vspace{0.2cm}

{\it The continuous hull.} The continuous hull $\Omega$ of the patterned waveguide is the topological space traced by the pattern when one continuously translates the waveguide along the axis \cite{Bel86}. Here we show that $\Omega$ is a 2-torus. With the elements introduced in diagram~\ref{Fig:Pattern}, let $\omega \in \TM^2$ be an arbitrary point and imagine the blue line as being a physical rope wound around the torus. Let the red line, whose equation is supplied in the diagram, be soaked with ink so that, every time when the rope crosses the red line, a mark is imprinted. Let us label these marks as shown in the diagram. Then, after we unwind the rope and lay it flat and parallel to axis of the tube, one will find that:
\begin{align}
x_{n+1} -x_n & = L_{\rm avg}+y\big ((n+1)\theta + \phi_\omega \big) -y(n\theta+\phi_\omega) \\ \nonumber
&= L_{\rm avg} + \Delta L \sin(n\theta + \phi_\omega).
\end{align} 
The marks $x_n$ will overlay perfectly over the centers of the walls if the origin of the Euclidean space is fixed at $x_\omega$ (the position of $\omega$ on the rope)! 

The conclusion is that every rigidly translated waveguide configuration can be uniquely characterized by a point $\omega \in \TM^2$, hence the continuous hull is the 2-torus. Furthermore, the group of translations parallel to the waveguide's axis induces an action $\tau$ of $\RM$ on $\TM^2$, which amounts to shifting $\omega$ along the winded rope. As such, the hull becomes a topological dynamical system $(\Omega,\tau,\RM)$.

\vspace{0.2cm}

{\it Algebra of physical observables.} When defining a topological invariant for aperiodic systems, the first task is to determine the operator algebra which supplies the associated physical observables. As it is now well known \cite{Bel86}, for continuous 1-dimensional models, this algebra is the crossed product $\Aa=C(\Omega) \rtimes_\tau \RM$. The elements of this algebra belong to a certain class of complex valued functions over $\RM \times \Omega$ and the multiplication rule is:
\begin{equation}
(f_1*f_2)(z,\omega)= \int_\RM d \xi \, f_1(\xi,\tau_{\xi-z}\omega) f_2(z-\xi,\omega).
\end{equation}
The algebra accepts a canonical representation on $L^2(\RM)$:
\begin{equation}
\big [ (\pi_\omega f) \psi \big ](z) = \int_\RM d \xi \,  f(z-\xi,\tau_{-\xi} \omega) \psi(\xi).
\end{equation}
Here, $z$ is the coordinate along the axis of the waveguide. The dispersion equation for our waveguide is defined over $L^2([0,R]) \otimes L^2(\RM)$, where $[0,R]$ is the interval where the radial coordinate $\rho$ takes values. The transversal modes, however, from topology point of view, brings nothing significant because all spectral projectors of the dispersion operator can be generated from the algebra $\KM \otimes \Aa$ (via the above representation), where $\KM$ is the algebra of compact operators over $L^2([0,R])$. This is the case because the resolvent of the radial part of the Laplace operator is compact when $\rho$ is restricted to a finite interval. 

\vspace{0.2cm}

{\it Topological Invariant.} We now can specify the input for the machinery developed in \cite{BourneMPAG2018}:
\begin{itemize}
\item If $(\omega_1,\omega_2)$ are the coordinates of $\omega \in \TM^2=\SM \times \SM$, then we have the derivations $\partial_{\omega_1}$ and $\partial_{\omega_2}$ ($=\partial_\phi$), as well as:
\begin{equation}
(\partial_z f)(z,\omega)= z f(z,\omega), \quad f \in \Aa.
\end{equation}
\item The trace ${\rm Tr} \otimes \Tt$ on $\KM \otimes \Aa$ with:
\begin{equation}
\Tt(f) = \int_{\TM^2} \frac{d\omega}{(2\pi)^2} \, f(0,\omega)
\end{equation}
\end{itemize}
Then, for a projection $p \in \KM \otimes \Aa$, \cite{BourneMPAG2018} showed that:
\begin{equation}
{\rm Ch}(p) = 2 \pi \, {\rm Tr}\otimes \Tt\big ( p [\partial_{\omega_2} p, \partial_z p]\big )
\end{equation}
equals the index of a certain Fredholm operator, which ensures the quantization and stability of this Chern number. We can use the physical representation to write this invariant. Indeed, if $P_\omega=\pi_\omega(p)$, then:
\begin{equation}\label{Eq:Chern}
{\rm Ch}(P) = \int d \rho \int_{\TM^2} d \omega \, \langle 0,\rho| P_\omega[\partial_\phi P_\omega,[Z,P_\omega]] |0,\rho \rangle,
\end{equation}
where $Z$ is the position operator parallel to the tube. Lastly, since $\tau$ acts ergodically on $\TM^2$, we have from Birkhoff theorem \cite{Bir}: 
\begin{equation}
\int d \rho \int_{\TM^2} d\omega \, \langle 0,\rho| \ldots |0,\rho \rangle = {\rm Tr}_L( \ldots ),
\end{equation}
the latter being  the trace per length introduced in the main text. With this simplification, Eq.~\eqref{Eq:Chern} becomes identical with the one supplied in the main text.

\vspace{0.2cm}

{\it Bulk-boundary correspondence.} According to \cite{BourneMPAG2018}, the topological class of $P_\omega$ is mapped into the $K_1$-class of the torus $\TM^2$ generated by the function $e^{\imath \phi}$. Since it involves only the vertical coordinate of the torus \ref{Fig:Pattern}, the horizontal coordinate plays no role in the bulk-boundary correspondence treated in our work.


\begin{thebibliography}{9}

\bibitem{ThoulessPRL1982} D. Thouless, M. Kohmoto, M. Nightingale, M. den Nijs, {\sl Quantized Hall Conductance in a Two-Dimensional Periodic Potential}, Phys. Rev. Lett. {\bf 49}, 405-409 (1982).

\bibitem{HaldanePRL1988} F.~D.~M.~Haldane, {\sl Model for a Quantum Hall-Effect without Landau levels: Condensed-matter realization of the parity anomaly}, Phys. Rev. Lett. {\bf 61}, 2015-2018 (1988).

\bibitem{SRFL2008} A.~P.~Schnyder, S.~Ryu, A.~Furusaki, A.~W.~W.~Ludwig, {\sl  Classification of topological insulators and superconductors in three  spatial dimensions}, Phys. Rev. {\bf B  78}, 195125 (2008).

\bibitem{QiPRB2008}  X.-L. Qi, T. L. Hughes, Shou-Cheng Zhang, {\sl Topological field theory of time-reversal invariant insulators}, Phys. Rev. B {\bf 78}, 195424 (2008).

\bibitem{Kit2009} A.~Kitaev, {\sl Periodic table for topological insulators and superconductors}, (Advances in Theoretical Physics: Landau Memorial Conference) AIP Conference Proceedings {\bf 1134}, 22-30 (2009).

\bibitem{RSFL2010}
S.~Ryu, A.~P. Schnyder, A.~Furusaki,  A.~W.~W. Ludwig, {\sl  Topological insulators and superconductors: tenfold way and  dimensional hierarchy}, New J. Phys. {\bf 12}, 065010 (2010).

\bibitem{HaldaneRaghu2008} F. D. M. Haldane, S. Raghu, {\sl Possible realization of directional optical waveguides in photonic crystals with broken time-reversal symmetry}, Phys. Rev. Lett. Lett. {\bf 100}, 013904 (2008).
 
 \bibitem{PP2009} E. Prodan, C. Prodan, 
{\sl Topological phonon modes and their role in dynamic instability of microtubules}, Phys. Rev. Lett. {\bf 103}, 248101 (2009).

\bibitem{WangNature2009} Z. Wang, Y. Chong, J. D. Joannopoulos, M. Soljacic, {\sl Observation of unidirectional backscattering-immune topological electromagnetic states}, 
Nature {\bf 461}, 772--775 (2009).

\bibitem{NashPNAS2015} L. M. Nash, D. Kleckner, A. Read, V. Vitelli, A. M. Turner, W. T. M. Irvine, 
{\sl Topological mechanics of gyroscopic metamaterials}, 
Proc. Nat. Acad. Sci. {\bf 112}, 14495-14500 (2015).

\bibitem{HafeziNatPhot2013} M. Hafezi, S. Mittal, J. Fan, A. Migdall, J. M. Taylor, {\sl Imaging topological edge states in silicon photonics}, Nature Photonics {\bf 7}, 1001-1005 (2013).

\bibitem{WuPRL2015} L.-H. Wu, X. Hu, {\sl Scheme for Achieving a Topological Photonic Crystal by Using Dielectric Material}, Phys. Rev. Lett. {\bf 114}, 223901 (2015).

\bibitem{SusstrunkScience2015} R. S{\"u}sstrunk, S. Huber, {\sl Observation of phononic helical edge states in a mechanical topological insulator}, Science  {\bf 349}, 47-50 (2015).

\bibitem{KaneNatPhys2013}
C. Kane, T. Lubensky, {\sl Topological boundary modes in isostatic lattices}, Nature Physics {\bf 10}, 39-45 (2013).

\bibitem{PauloseNatPhys2015} J. Paulose, B. G. Chen, V. Vitelli, {\sl Topological modes bound to dislocations in mechanical metamaterials}, Nature Physics {\bf 11}, 153-156 (2015).

\bibitem{ProdanNatComm2017} E. Prodan, K. Dobiszewski, A. Kanwal, J. Palmieri, C. Prodan, {\sl Dynamical Majorana edge modes in a broad class of topological mechanical systems}, Nature Communications {\bf 8}, 14587 (2017).

\bibitem{KhanikaevNatPhot2017} A. Slobozhanyuk, S. H. Mousavi, X. Ni, D. Smirnova, Y. S. Kivshar, A. B. Khanikaev, {\sl Three-dimensional all-dielectric photonic topological insulator}, Nature Photonics {\bf 11}, 130-136 (2017).

\bibitem{MousaviNatComm2015} S. H. Mousavi, A. B. Khanikaev, Z. Wang, {\sl Topologically protected elastic waves in phononic metamaterials}, Nature Communications {\bf 6}, 8682 (2015).

\bibitem{RuzzeneArxiv2017} M. Miniaci, R. K. Pal, B. Morvan, M. Ruzzene, {\sl Observation of topologically protected helical edge modes in Kagome elastic plates}, arXiv:1710.11556 (2017).
 
\bibitem{ChaunsaliPRB2018} R. Chaunsali, C.-W. Chen, J. Yang, {\sl Subwavelength and directional control of flexural waves in zone-folding induced topological plates}, Phys. Rev. B {\bf 97}, 054307 (2018).
 
\bibitem{ChernArxiv2018} H. Chen, H. Nassar, G. Huang, {\sl Topological mechanics of edge waves in Kagome lattices}, arXiv:1802.04404 (2018). 

\bibitem{PalJAP2016} R. K. Pal, M. Schaeffer, M. Ruzzene, {\sl Helical edge states and topological phase transitions in phononic systems using bilayered lattices}, J. Appl. Phys. {\bf 119}, 084305 (2016).


\bibitem{KLR2012} Y. E. Kraus, Y. Lahini, Z. Ringel, M. Verbin, O. Zilberberg, {\sl Topological states and adiabatic pumping in quasicrystals}, Phys. Rev. Lett. {\bf 109}, 106402 (2012).

\bibitem{VZK2013} M. Verbin, O. Zilberberg, Y. E. Kraus, Y. Lahini, Y. Silberberg, {\sl Observation of topological phase transitions in photonic quasicrystals}, Phys. Rev. Lett. {\bf 110}, 076403 (2013).

\bibitem{ProdanPRB2015} E. Prodan, {\sl Virtual topological insulators with real quantized physics}, Phys. Rev. B {\bf 91}, 245104 (2015).

\bibitem{BabouxPRB2017} F. Baboux, E. Levy, A. Lemaitre, C. Gomez, E. Galopin, L. Le Gratiet, I. Sagnes, A. Amo, J. Bloch, E. Akkermans, {\sl Measuring topological invariants from generalized edge states in polaritonic quasicrystals}, Phys. Rev. B \textbf{95}, 161114 (2017). 

\bibitem{Apigo2} D.J. Apigo, K. Qian, C. Prodan, and E. Prodan, {\sl Topological edge modes by smart patterning}, Phys. Rev. Materials {\bf 2}, 124203 (2018).

\bibitem{Bel86} J.\ Bellissard, {\sl K-theory of $C^*$-Algebras in solid state physics}, Statistical mechanics and field theory: mathematical aspects, (Springer, Berlin, Heidelberg, 1986, 99-156).

\bibitem{PS} E.~Prodan, H.~Schulz-Baldes, {\sl Bulk and boundary invariants for complex topological insulators: From $K$-theory to physics}, (Springer, Berlin, 2016).

\bibitem{Xiao} M. Xiao, G. Ma, Z. Yang, P. Sheng, Z.Q. Zhang, C.T. Chan, {\sl Geometric phase and band inversion in periodic acoustic systems}, Nature Physics, {\bf 11}:240, (2015).

\bibitem{He} C. He, X. Ni, H. Ge, X. Sun, Y. Chen, M. Lu, X. Liu, Y. Chen, {\sl Acoustic topological insulator and robust one-way sound transport}, Nature Physics, {\bf 12}:1124, (2016).

\bibitem{NiArxiv2018} X. Ni, M. Weiner, A. Alù, A. B. Khanikaev, {\sl Observation of bulk polarization transitions and higher-order embedded topological eigenstates for sound}, arXiv:1807.00896 (2018).

\bibitem{RichouxEPL2002} O. Richoux, V. Pagneux, {\sl Acoustic characterization of the Hofstadter butterfly with resonant scatterers}, EPL (Europhysics Letters), Vol 59, 34 (2002).

\bibitem{BourneMPAG2018} C. Bourne, A. Rennie, {\sl Chern numbers, localisation and the bulk-edge correspondence for continuous models of topological phases}, Math. Phys. Anal. Geom. 21: 16. https://doi.org/10.1007/s11040-018-9274-4 (2018).

\bibitem{ProdanSpringer2017} E. Prodan, {\sl A computational non-commutative geometry program for disordered topological insulators}, Springer Briefs in Mathematical Physics, Springer, 2017.

\bibitem{ProdanJPA2018}  C. Bourne, E. Prodan, {\sl Non-commutative Chern numbers for generic aperiodic discrete systems}, J. Phys. A: Math. Theor. {\bf 51}, 235202 (2018).

\bibitem{Suppl} See Supplemental Material at [URL] for a computation of the continuous hull of the pattern as well as for the definition of the bulk topological invariant. 

\bibitem{HofstadterPRB1976} D.~R.~Hofstadter, {\sl Energy levels and wave functions of Bloch electrons in rational and irrational magnetic fields}, Phys. Rev. B {\bf 14}, 2239-2249 (1976).

\bibitem{Bir} G.~D.~Birkhoff, {\sl  Proof of the ergodic theorem}, Proc. Natl. Acad. Sci. USA {\bf 17}, 656-660 (1931).

\bibitem{ProdanJGP2018} E. Prodan, Y. Shmalo, {\sl The K-Theoretic Bulk-Boundary Principle for Dynamically Patterned Resonators}, Journal of Geometry and Physics {\bf 135}, 135-171 (2019).


\end{thebibliography}
\end{document}